\documentstyle[prb,epsfig,aps]{revtex}

\begin{document}
\title{Thermal equation of state of tantalum}
\author{Ronald E. Cohen$^{(1,2)}$ and O. G\"{u}lseren$^{(1,3,4)}$}
\address{$^{(1)}$Geophysical Laboratory and Center for High Pressure Research, 
Carnegie Institution of Washington, 5251 Broad Branch Road, NW, Washington, DC 20015}
\address{$^{(2)}$Seismological Laboratory, California Institute of
Technology, Pasadena, CA 91125}
\address{$^{(3)}$NIST Center for Neutron Research, Gaithersburg, MD 20899 }
\address{$^{(4)}$Department of Materials Science,
University of Pennsylvania, Philadelphia, PA 19104}
\date{\today}
\maketitle

\begin{abstract}
We have investigated the thermal equation of state of bcc tantalum from
first principles using the full-potential Linearized Augmented Plane Wave (LAPW) and
mixed-basis pseudopotential methods for pressures up to 300 GPa and temperatures up to 10000 K. The equation of state at zero temperature was computed using LAPW. For finite temperatures, mixed basis
pseudopotential computations were performed for 54 atom supercells. The vibrational contributions were obtained by computing the partition function
using the particle in a cell model, and the the finite temperature
electronic free energy was obtained from the LAPW band structures. We discuss the behavior of thermal equation of state parameters such as the
Gr\"uneisen parameter $\gamma$, the thermal expansivity $\alpha$, the Anderson-Gr\"uneisen parameter $\delta _T$ as
functions of pressure and temperature. The calculated Hugoniot shows excellent agreement with shock-wave experiments.
An electronic topological transition was found at approximately 200 GPa.
\end{abstract}

\pacs{PACS numbers: 05.70.Ce, 64.30.+t, 71.20.Be, 65.50.+m, 65.70.+y}

\narrowtext

\section{Introduction}

We investigate from first-principles the thermal equation of state of
body-centered cubic (bcc) tantalum, a group V transition metal, which is a useful 
high-pressure standard due to its high
structural mechanical, thermal and chemical stability. Ta has a very high
melting temperature, 3269~K at ambient pressure, and its bcc structure is
stable for a large pressure range. Static diamond anvil cell experiments~\cite{cynn} up to 174~GPa
and full-potential linearized muffin
tin orbital (LMTO) calculations~\cite{soderlind} up to 1~TPa conclude bcc
phase of Ta is stable at these pressure range. Similarly, shock compression
experiments\cite{hug1} showed melting at around 300~GPa, but no solid-solid
phase transition.

\section{Static equation of state}

\label{sec:static}

\begin{figure}[tbp]
\centerline{\epsfig{file=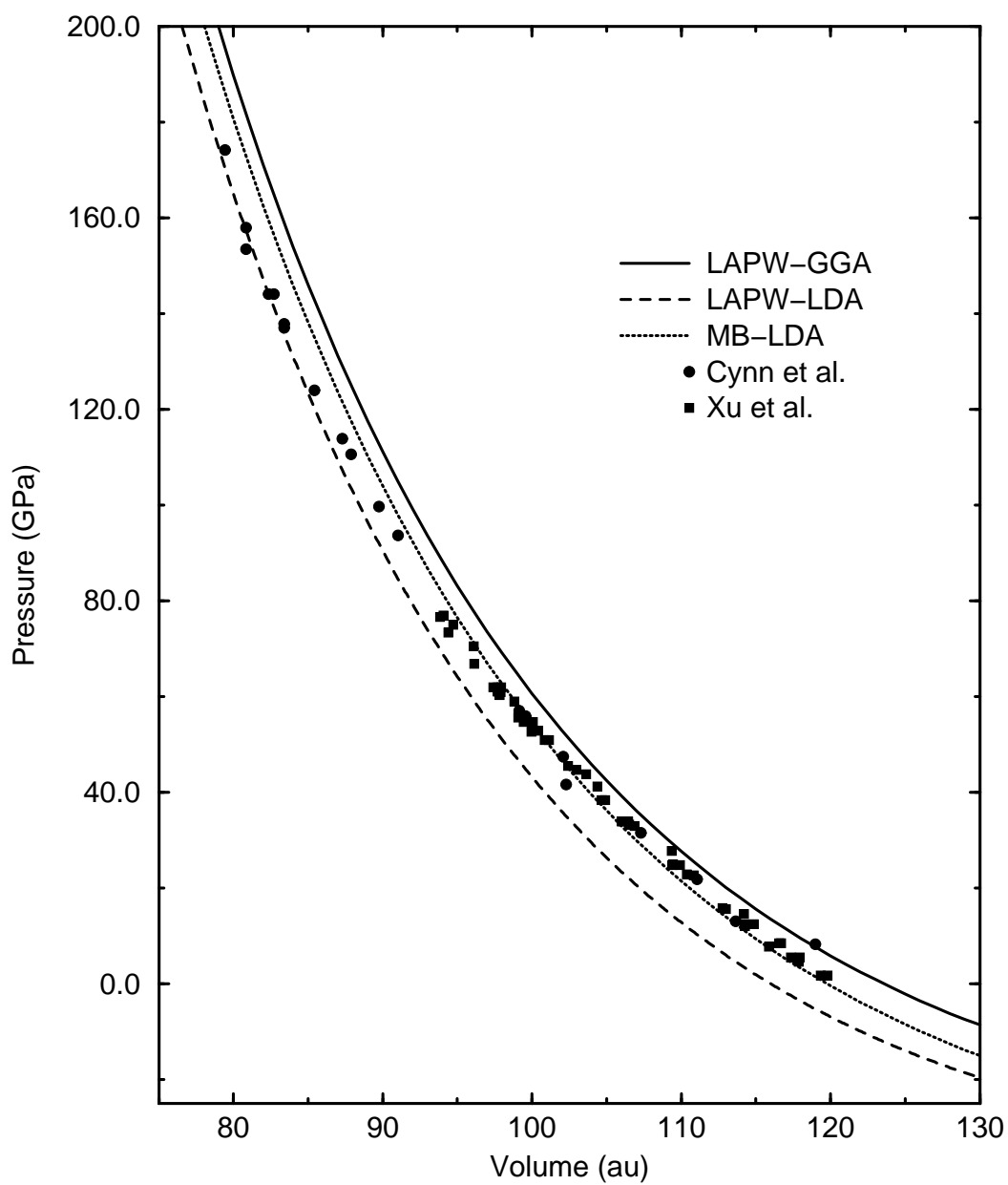,height=7in}}
\vspace{10pt}
\caption{Static equation of state of Ta. Solid and dashed lines are LAPW GGA
and LDA calculations respectively. Pseudopotential mixed-basis results are
shown by the dotted line. Circles and squares are two different diamond
anvil cell experiments. The discrepancies at high pressures may be due to
strength effects in the static equation of state.}
\label{fig:eost0K}
\end{figure}

Firstly, we discuss the static high pressure properties of Ta, which we
obtained from first principles by using the LAPW method \cite%
{weikrakauer,singh}. The $5p$,$4f$,$5d$ and $6s$ states were treated as band
states, others are described as core electrons. We used both the local
density approximation (LDA) \cite{hedin} and generalized gradient
approximation (GGA) \cite{pbe} for the exchange-correlation potential. The
Monkhorst-Pack special $\vec{k}$-point scheme~\cite{monpack} with a 16x16x16
k-point mesh (140 k-points within the irreducible Brillouin zone of the bcc
lattice) was used after convergence tests. The convergence parameter RK$%
_{max}$ was 9.0, and the muffin tin radii were 2.0 bohr, giving about 1800
plane-waves and 200 basis functions per atom at zero pressure. The total
energy was computed for 20 different volumes from 62.5 to 164 bohr$^{3}$ (1
bohr=0.529177 $\text{\AA}$), and the energies were fit to the Vinet equation~\cite{vinet}, 
\begin{eqnarray}
E(V,T) &=&E_{0}(T)+\frac{9K_{0}(T)V_{0}(T)}{\xi ^{2}}\{1+\{\xi (1-x)-1\}
\label{vinetev} \\
&&\times exp\{\xi (1-x) \nonumber \}\}
\end{eqnarray}
where $E_{0}$ and $V_{0}$ are the zero pressure equilibrium energy and volume respectively, $x=(\frac{V}{V_{0}})^{\frac{1}{3}}$ and $\xi =\frac{3}{2}(K_{0}^{\prime }-1)$
, $K_{0}(T)$ is the bulk modulus and $K_{0}^{\prime }(T)=(\partial
K(T)/\partial P)_{0}$.
The subscript 0 alone throughout represents the standard state P=0. All equations of state here are for an
isotherm or static (T=0) conditions, unless specified otherwise. Pressures were obtained analytically from 
\begin{equation}
P(T,V)=\{\frac{3K_{0}(T)(1-x)}{x^{2}}\}exp\{\xi (1-x)\}.  \label{vinetpv}
\end{equation}
The calculated equation of state is compared with experiments~\cite{cynn,eosmao} in Fig.~\ref{fig:eost0K}. The LAPW GGA results are found to be
more accurate than the LDA. The discrepancies are larger beween theory and experiment at high pressures; this may be due to strength effects in the experiments~\cite{duffy}, since we find good agreement with the experimental Hugoniot to 400 GPa (discussed below). For the pseudopotential mixed-basis calculations~\cite{mixed,oguz} (MBPP, discussed below), we find that the LDA agrees fairly well with the experiments, indicating compensating errors between the pseudopotential and the LDA. Since we are using a first-principles approach, and want to avoid ad hoc variations in procedure to get better agreement with experiments, we use the computational more accurate method with the least approximations, that is LAPW and GGA, rather than the MBPP with LDA, in spite of the fortuitously better agreement of the latter with the room temperature data. 

\begin{figure}[tbp]
\centerline{\epsfig{file=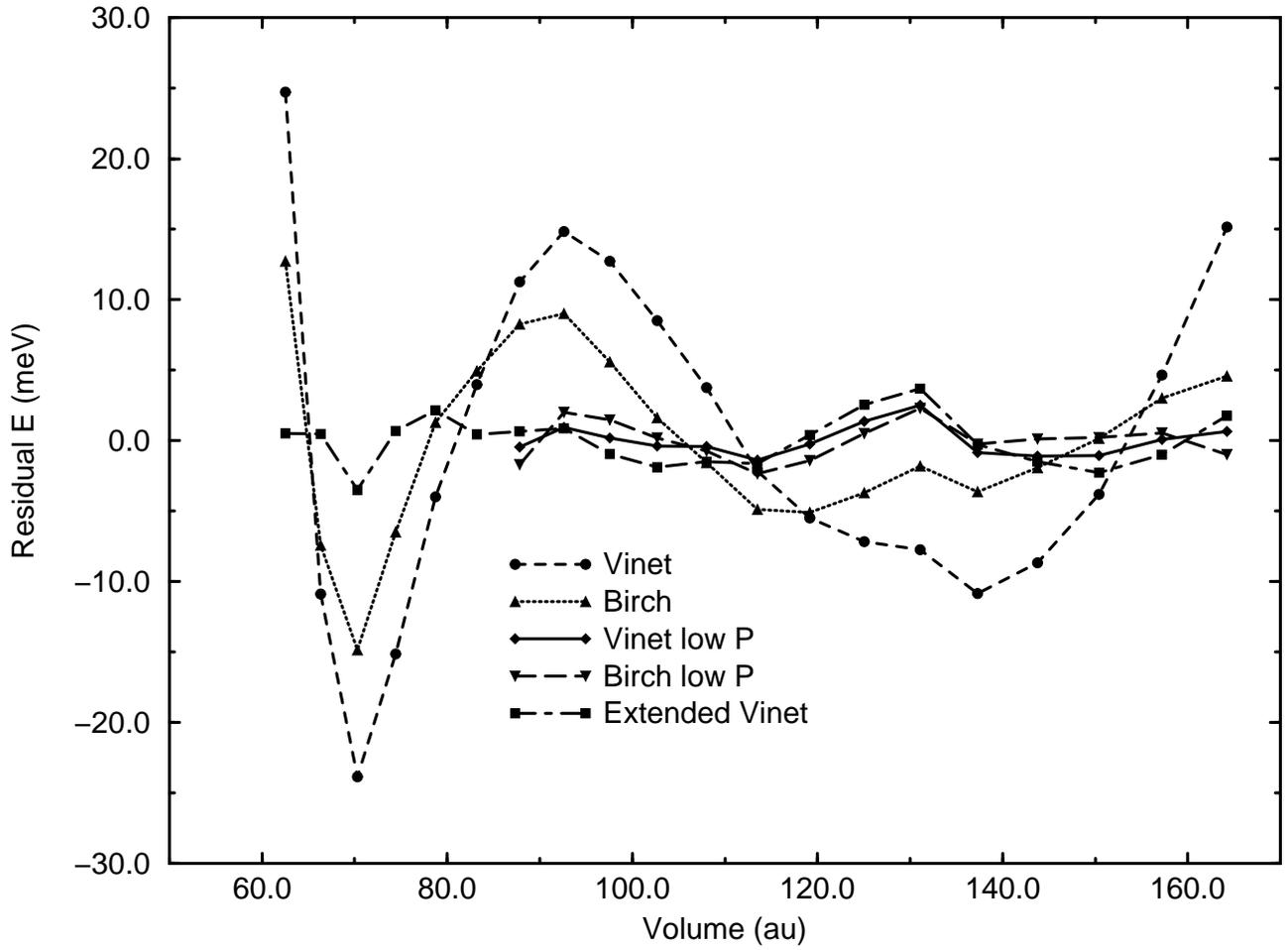,height=5in}}
\vspace{10pt} 
\caption{Energy differences between the calculated and fitted data of the
equation of state fits. Both Vinet and Birch-Murnaghan equation-of-state
fits are shown. In the low pressure fits only the data with volume greater
than 84~bohr$^3$ were included. The extended Vinet fit included the $2^{nd}$
and $3^{th}$ pressure derivative of $K_0$(T).}
\label{fig:res}
\end{figure}

\begin{figure}[tbp]
\centerline{\epsfig{file=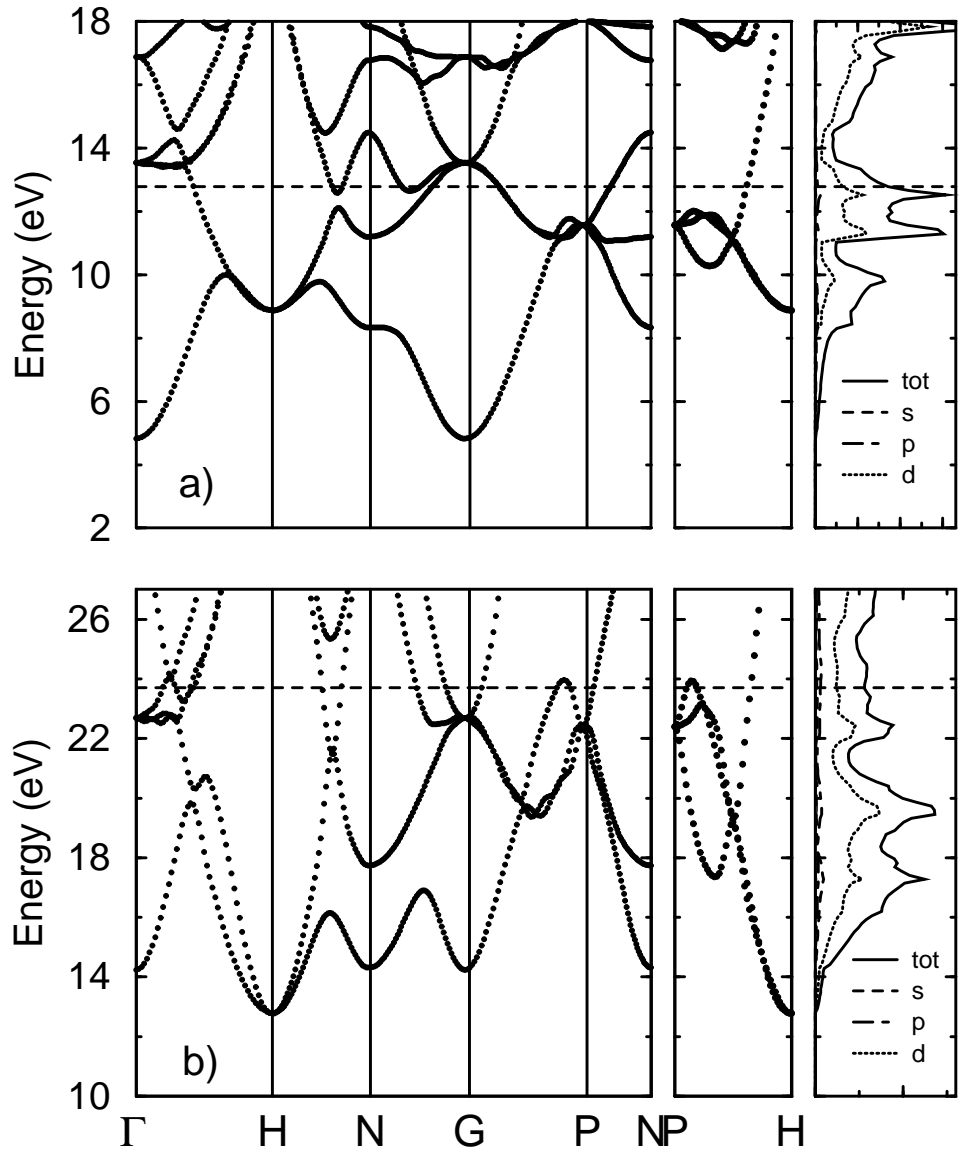,height=6in}}
\vspace{10pt} 
\caption{Band structures and densities of states at two different volumes
(pressures): (a) V=120~bohr$^3$ (P=5~GPa) (b) V=50~bohr$^3$. (P=460~GPa).
Note the significant change in Fermi surface configuration with pressure.}
\label{fig:banddos}
\end{figure}

The residuals between the calculated and fitted energies show large
deviations for volumes less than 80~bohr$^3$ (Fig.~\ref{fig:res}). Note that
the residuals are all small and not evident for the large energy scale shown in Fig.~\ref{fig:globalf}a. Other equation of state
formulations, such as the extended Birch equation \cite{birch} show the same trend~\cite{ron1}. When the fit is restricted to
volumes greater than 80~bohr$^3$, or extending the Vinet equation by two
more parameters related to the next two pressure derivative of bulk modulus,
it is improved significantly. Hence, these large residuals are related to
different high- and low-pressure behavior of Ta. The band structures and
densities of states (Fig.~\ref{fig:banddos}) show a major reconfiguration of
the Fermi surface. This electronic topological transition is the reason for
this change in compression, and the behavior of the residuals for the fitted
equations of state. This indicates that systematic deviations from simple equations of state can
be used to find subtle phase transitions \cite{ron1}.

Spin-orbit interactions may be important for Ta. In order to test this, we
included spin-orbit coupling by second variational treatment~\cite%
{2variation} including 20-80 bands. In contrast to the fully relativistic
LMTO results~\cite{soderlind}, we found only negligible effect on the
equation of state, so our computations were done without spin-orbit.

\section{Calculation of Thermal Properties}

In order to compute the high temperature properties of Ta, we separated the
Helmholtz free energy as~\cite{wass}: 
\begin{equation}
F(V,T) = E_{static}(V) + F_{el}(V,T) + F_{vib}(V,T) ,
\end{equation}
where $E_{static}(V)$ is the static zero temperature energy, $F_{el}(V,T)$
is the thermal free energy from electronic excitations, and $F_{vib}(V,T)$ is the vibrational
contribution to the free energy. $E_{static}(V)$ and $F_{el}(V,T)$ were
computed using the LAPW method with the GGA. $F_{vib}(V,T)$ was computed using the
particle-in-a-cell (PIC) model with a mixed basis pseudopotential method, as described below. The main differences we
find between LDA and GGA are in the energy versus compression, but we find only small differences in energy versus
atomic displacements. Since LDA converges much faster than GGA, and is also faster per iteration cycle, we used LDA for the large
supercell computations required for the vibrational contributions. Differences from using GGA for this are
negligible.

The electronic part $F_{el}(V,T)$ of the free energy is: 
\begin{equation}
F_{el}(V,T) = E_{el}(V,T) - TS_{el}(V,T)
\end{equation}
where $E_{el}(V,T)$ is the internal energy due to thermal electronic
excitations,
\begin{equation}
S_{el}(V,T)=-2k_{B}\sum_{i}f_i\ln f_i+(1-f_i)\ln (1-f_i)  \label{mermin}
\end{equation}
is the electronic entropy, and the Fermi-Dirac occupation $f_i$ is 
\begin{equation}
f_i = \frac{1}{1+\exp(\frac{(\epsilon_i - \mu(T))}{k_BT})},
\end{equation}
$\epsilon_i$ are the eigenvalues, $\mu$ is the chemical potential, and $k_B$
is the Boltzmann constant. The vibrational free energy is given in terms of
the partition function as 
\begin{equation}
F_{vib}(V,T) = -k_BT\ln Z.  \label{vibfree}
\end{equation}

The particle-in-a-cell model \cite{wass,cellmodel} was used to calculate the
partition function. In this model, the partition function is factored by
neglecting atomic correlations. An atom is displaced in its Wigner-Seitz cell in the potential field of all the other
atoms fixed at their equilibrium positions, i.e. the ideal, static lattice
except for the wanderer atom. The partition function is simply a product of
identical functions for all the atoms, involving an integral of Boltzmann
factor over the position of a single atom inside the Wigner-Seitz cell, 
\begin{equation}
Z_{cell} = \lambda^{-3N} \{ \int \exp\left[-\frac{(U(\vec{r})-U_0(T_0))}{k_BT%
}\right]d\vec{r}\}^{N},  \label{cellpart}
\end{equation}
where $\lambda=h/(2\pi mk_BT)^{1/2}$ is the de Broglie wavelengths of atoms
and $U(\vec{r})$ is the potential energy of the system with the wanderer atom
displaced by radius vector $\vec{r}$ from its equilibrium position. The
advantage of the cell model over lattice dynamics based on the quasiharmonic
approximation is that anharmonic contributions from the potential-energy of
the system have been included exactly without a perturbation expansion. On
the other hand, since the interatomic correlations between the motions of
different atoms is ignored, it is only valid at temperatures above the Debye
temperature. Diffusion and vacancy formation are also ignored, so premelting
effects are not included. We have used the classical partition function, so quantum phonon effects are not included. Thus the
heat capacity and thermal expansivity do not vanish at low temperatures, for example. The present results are appropriate
for temperatures above the Debye temperature (245 K in Ta \cite{grigoriev}) and below premelting effects.
Since the Debye temperature is below room temperature in Ta, the classical thermal properties should be reasonable
even down to room temperature.

The electronic thermal free energy was obtained using the Mermin theorem~%
\cite{temdft} (Eq.~\ref{mermin}). The charge density is temperature
dependent through both occupation numbers according to the Fermi-Dirac
distribution and self-consistency. The electronic contributions to the
thermal free energy were computed by the LAPW method using the same
computational parameters as the $T=0 K$ computations described in Section~\ref{sec:static}.

For the vibrational contributions, it is necessary to do a large number of
large supercell calculations, which is computationally intractable by the
LAPW method, but is achievable with the MBPP method~\cite{oguz}. In this mixed-basis approach, pseudo-atomic
orbitals and a few low-energy plane waves are used as the basis set within a
density functional, pseudopotential calculation. It was shown that the
method offers a computationally efficient but accurate alternative.

A semi-relativistic, nonlocal and norm-conserving Troullier-Martins~\cite%
{tromartins} pseudopotential (with associated pseudo-atomic orbitals) was
used to describe the Ta atoms. The pseudopotential was generated from an $%
5d^36s^26p^0$ atomic configuration with cutoff radii 1.46, 2.6 and 3.4 bohr
for $5d$, $6s$, and $6p$ potentials, respectively, with non-linear core
corrections. The cutoff radii were optimized by testing the transferability
of the pseudopotential by considering the reasonable variations of the
reference atomic configuration and by comparing the logarithmic derivative
of the pseudo-wavefunctions with all-electron values in the valence energy
range. The $6s$ potential was chosen as the local component while $5d$ and $%
6p$ were kept as nonlocal while transforming the potential to the nonlocal
separable Kleinman-Bylander form~\cite{kleinman}. A full plane-wave
representation is used for the charge density and potential, and a smaller cutoff is used for the basis set. The pseudoatomic
orbitals are expanded in the large plane wave set for evaluation of the potential and charge density integrals in the
Hamiltonian and overlap matrices, and in the total energies. After checking the energy convergence, 550~eV and 60~eV
were used for the large and small energy cutoffs, respectively, in the solid
calculations. The exchange-correlation effects of electrons were treated
within LDA. The $T=0K$ equation of state of bcc Ta was computed to test the
pseudopotential, and is compared with experiment and LAPW results in Fig.~\ref{fig:eost0K}.

For the PIC computations, a supercell with 54 atoms was used. The
MBPP calculations were carried out on this 54 atoms supercell using LDA for exchange-correlations
effects and 4 special $\vec{k}$ points for BZ integrations. The potential
energy surface was then calculated as a function of the displacements of the
wanderer atom. Symmetry was taken into account in order to reduce the number
of computations. The integrand in Eq.~\ref{cellpart} has a
Gaussian-like shape and decays rapidly, and essentially is zero at half of
the interatomic distances even at very high temperatures. Therefore,
integration over the Wigner-Seitz cell can be replaced by an integration
over the inscribed sphere. Also, the radial part of the integrand is
invariant under point group operations of the lattice, hence a numerical
quadrature can be used for angular integration based on the method of
special directions~\cite{wass,spdir}. In this method, the radial integral is
expanded in terms of lattice harmonics, cubic harmonics for a cubic lattice,
then a quadrature rule is derived for the angular integration in terms of
the radial integration by choosing special directions, $\hat{r}_i$, in such
a way that the contribution from $l \ne 0$ terms, as many lattice harmonics
as possible, is zero. In all of the computations, we used one special
direction which integrates exactly up to $l=6$ cubic harmonics~\cite{spdir}.
Then, the potential energy was calculated at 4-6 different displacements
along this special direction. In order to model the potential, these
computed values were fit to an even polynomial up to order 8, which shows
the anharmonicity very clearly, since a second order fit describes the data
poorly. Finally, the cell-model partition was calculated from Eq.~\ref%
{cellpart} by carrying out the integration numerically, and the vibrational
free energy is simply given in terms of partition function by Eq.~\ref%
{vibfree}.

\section{Thermal equation of state}

We treated the resulting free energies \cite{web} three different ways.
Firstly, F-V isotherms were fit using the Vinet equation of state, giving $E_0(T)$, $V_0(T)$, $K_0(T)$ and $K_0^{\prime}(T)$
as the parameters of the fit (Table \ref{table:vinet}, Fig.~\ref%
{fig:vinetvke}). Because of the thermal expansivity, the minimum energy
shifts to higher volumes with increasing temperature, and above T=6000~K the
P=0 volume is not in the range of volumes we studied. The experimental P=0
melting temperature is 3270 K, so temperatures above this are non-physical in any case. The parameters for higher
temperatures are fictive parameters that describe the higher pressure equation of state accurately. The Vinet parameters could
be fit to polynomials versus temperature to obtain a thermal equation of state, but the following approaches require
fewer parameters.

A second thermal equation of state was obtained by analyzing the thermal
pressure obtained from the Vinet fits (i.e.~the differences in pressures between isotherms). The thermal pressure
as a function of volume and temperature is shown in Fig.~\ref{fig:pthermal}. The volume dependence of the thermal
pressure for Ta is very weak up to 80\% compression. Deviations at higher compressions may be due to the fit through
the electronic topological transition discussed above, and thus due to inflexibility in the Vinet equation, rather
than a real rise in thermal pressure. The thermal pressure is also quite linear in T. The thermal pressure changes are
given by 
\begin{equation}
P(V,T) - P(V,T_0)=\int_{T_0}^T \alpha K_T dT
\end{equation}
and $\alpha K_T$ is quite constant for many materials \cite{anderson} in the classical regime (above the Debye temperature). Hence, a greatly simplified thermal equation of state is to add a thermal
pressure 
\begin{equation}
P_{th}(T) = a T  \label{eq:simpleeos}
\end{equation}
to the static pressure $P_{static}$. Since our equation of state is classical, we can use this expression at all temperatures. This simple equation of state gives $q=\left(\frac{\partial \ln \gamma}{\partial \ln V}\right)_T=1$ and $\left(\frac{\partial K_T}{\partial T}\right)_V=0,$ which therefore are good
approximations over this pressure and temperature range for Ta.

The thermal pressure was averaged over volumes from 60 bohr$^3$ to 220 bohr$%
^3$, and is shown as a function of temperature in Fig.~\ref{fig:pavth}. The solid line has a slope $a$ of
0.00442 GPa/K. This is close to $\alpha K_{0T}(T)$, which is 0.00460 GPa/K
at 1000 K and zero pressure. So a simple equation of state for Ta is the 
static pressure given by V$_0$(T=0)=123.632 bohr$^3$, K$_0$(T=0)=190.95 GPa,
and K$_0^\prime$(T=0)=3.98 in the Vinet equation (Eq.~\ref{vinetpv}) plus the thermal pressure P$_{th} = 0.00442 T$.

Thirdly, an accurate high temperature global equation of state was formed
from the $T=0K$ Vinet isotherm and a volume dependent thermal free energy $%
F_{th}$ as: 
\begin{equation}
F_{th} = \sum_{i=1,j=0}^{i=3,j=3} A_{ij} T^i V^j - 3 k_B T \ln T
\label{eq:globaleos}
\end{equation}
This is the thermal Helmholtz free energy per atom, which must be added to
equation \ref{vinetev} to obtain the total free energy. The parameters $A_{ij}$ are given in SI units in Table \ref%
{table:params}, which gives the free energy in Joules/atom; T is in Kelvin and V in m$^3$/atom. The term in $T \ln T$ is
necessary to give the proper classical behavior at low temperatures, since we are evaluating the classical
partition function, with $C_V=3 k_B$ and $S=-\infty$ at $T=0 K$. For the best overall accuracy, the T=0 isotherm was
also included in the global fit. The global fit is compared with the computed free energies in Fig.~\ref%
{fig:globalf}. The r.m.s. deviation of the fit is 0.4 mRy. At low temperatures (0 and 1000 K) the residuals (Fig.~\ref%
{fig:globalf}b) are larger due to the electronic topological transition discussed above (Fig.~\ref{fig:res}), but at higher
temperatures this anomaly is less pronounced due to the thermal smearing, and the equation of state fits well.

Thermal equation of state parameters such as $P$,$\alpha$,$\gamma$,$\delta _T
$,$q$, and the heat capacity $C_V$ and $C_P$ can be obtained from the global fit by differentiation and algebraic
manipulation (see \cite{anderson} for a collection of useful formulas.) We now discuss the behavior of these
parameters. The thermal expansion coefficient is presented, and compared with zero pressure experiments \cite{toul} in Fig.~%
\ref{fig:alpha}. The deviations at lower temperatures are due to the use of
the classical partition function. The thermal expansivity is a quite sensitive parameter, and the errors at
moderate temperatures, which are typical in first-principles computations, may come from a number of sources (error in
the P=0 volume, LDA, the pseudopotential, the PIC model, or convergence in k-points or basis set). The divergence in
behavior at higher temperatures is not due to vacancy formation, since the vacancy formation in Ta is high (3.2 eV) %
\cite{vacancy}, and the fraction of vacancies at the melting point is less than 10$^{-4}$. The temperature range over
which the anomaly occurs seems too large to be a premelting effect. One possible explanation would be an incipient
solid-solid phase transition in Ta, which would not be detected in the PIC method. The upturn in $\alpha$ with increasing
temperature is apparently a low pressure phenomenon. It is possible that it is due to an experimental problem, such
as oxidation of the sample.

The thermal expansivity drops rapidly with increasing pressure (Fig.~\ref%
{fig:alpha}), and this is parametrized by the Anderson-Gr\"{u}neisen parameter (Fig.~\ref{fig:deltaT}) 
\begin{equation}
\delta _T=\left(\frac{\partial \ln \alpha}{\partial \ln V}\right)_T.
\end{equation}
The behavior of $\delta _T$ is complex. At low pressures it increases with
increasing temperature, but at elevated pressures it decreases with temperature. The parameter $\delta _T$ can be
fit to a form~\cite{andisaak} $\delta _T=\delta _T(\eta=1) \eta^\kappa$
where $\eta=V/V_0(T_0)$. The average $\delta _T$ (averaged from 0-6000 K) 
decreases with compression, and a power law fit gives $\delta
_T(\eta)=4.56\eta^{1.29}$; at 1000 K $\delta _T=4.75\eta^{1.17}$. Interestingly these values are not that different from
MgO~\cite{andisaak}  ($\delta _T(\eta=1, 1000K))=5.00$ and $\kappa=1.48$.
The behavior is much different than for Fe, where $\delta _T$ is constant to 150 GPa with values of 5.2 and 5.0 for fcc and hcp,
respectively, after which it drops more slowly than a power law \cite{wass}. The difference between $\delta _T$ and $K^\prime$ is
an important anharmonic parameter, and is related to the change in the bulk modulus with temperature at constant
volume and the thermal pressure with compression at constant temperature: 
\begin{eqnarray}
\delta _T-K^\prime & = & \left(\frac{\partial \ln (\alpha K_T)}{\partial \ln
V}\right)_T \\
& = & \frac{-1}{\alpha K_T} \left(\frac{\partial K_T}{\partial T}\right)_V.
\end{eqnarray}
Fig.~\ref{fig:deltaT}d shows that $\delta _T-K^\prime$ is quite small over a
large temperature and compression range, but increases at high T and P. This is consistent with the accuracy of the
simple equation of state (Eq.~\ref{eq:simpleeos}). The behavior of $\delta _T-K^\prime$ is also
surprisingly similar to the behavior of MgO (see Fig.~3.3 in \onlinecite{anderson}).

Changes in thermal pressure $P_{th}$ are given by $\alpha K_T=\left(\frac{%
\partial P}{\partial T}\right)_V$ which is shown in Fig.~\ref{fig:akt}.
Changes in $\alpha K_T$ are small, but it is interesting that the sign of
the change with temperature is strongly dependent on pressure, indicating that
experiments to determine $\left( \frac{\partial\alpha K_T}{\partial T}\right)
$ at low pressures may not be applicable to a very large pressure range.

A most important parameter, particularly for reduction of shock data, is the
Gr\"{u}neisen parameter 
\begin{equation}
\gamma=V\left(\frac{\partial P}{\partial E}\right)_V=\frac{\alpha K_T V}{C_V}%
,  \label{eq:gamma}
\end{equation}
where $E$ is the internal energy. The Gr\"{u}neisen parameter is used in the
Mie-Gr\"{u}neisen equation of state, which assumes $\gamma$ independent of temperature. Then the thermal pressure
on the Hugoniot, for example is given by the change in internal energy by 
\begin{equation}
P_{hug}-P_{static}=\frac{\gamma}{V}(E_{hug}-E_{static}).
\end{equation}
Fig.~\ref{fig:gamma} shows that at elevated pressures, $\gamma$ is
moderately temperature dependent, and it varies more strongly with temperature below 100 GPa. The variation of $\gamma$ with
pressure is given by 
\begin{equation}
q=\frac{\partial \ln \gamma}{\partial \ln V}
\end{equation}
which is shown in Fig.~\ref{fig:q}. The parameter $q$ is not constant, as is
often assumed, but decreases significantly with pressure and temperature. If $\alpha K_T$ and $C_V$ were
constant, Eq.~\ref{eq:gamma} shows that $q=1$. Fig.~\ref{fig:akt} shows that $\alpha K_T$ is quite constant, so that
large changes in $q$ must be due primarily to changes in the heat capacity $C_V$.

Fig.~\ref{fig:heatcapacity} shows indeed that the heat capacity is a strong
function of temperature and pressure. This is due mainly to the electronic contributions. The experimental C$_P$ at
zero pressure~\cite{grigoriev} is also shown. Other than the large differences from experiment at very low temperatures,
due to neglect of quantum phonon effects in the present model, there is a large increase in the experimental heat
capacity with increasing temperature that is not seen in the PIC results. A similar large increase in the experimental
thermal expansivity is not predicted by the model (Fig.~\ref{fig:alpha}). Vacancy formation (not included in the PIC
model) seems an unlikely source for this discrepancy as discussed above. An incipient phase transition or sample
oxidation seems the most likely cause of the observed behavior in the thermal expansivity and the heat capacity.

To compare with experiment at high pressures and temperatures, we consider
the high temperature, high pressure equation of state obtained by shock compression \cite{hug1}. The pressures, $%
P_H$ and temperatures, $T_H$, on the Hugoniot of Ta are given by the
Rankine-Hugoniot equation: 
\begin{equation}
\frac{1}{2} P_H(V_0(T_0) - V) = E_H - E_0(T=0) .
\end{equation}
We solved the Rankine-Hugoniot equation using our equation of state results
by varying the temperature at a given volume until it is satisfied. The calculated Hugoniot shows very good
agreement with experimental data as seen in Fig.~\ref{fig:hug}.

Computations of the Hugoniot using a modified free-volume method were
recently presented by Wang {\it et al.}~ \cite{wang}. Though superficially similar to the PIC method, their model is
approximate. They included the electronic free energy in the same way that we do, but no supercell is used for the
phonon contribution. Instead they find an effective potential from the equation of state of the primitive, one atom
unit cell, and integrate the mean field phonon partition function based on this effective potential. This is a
tremendous reduction in effort compared with the use of large supercells and finding the potential for displacing one
atom in the supercell. They obtain impressive results for this simple model obtaining excellent agreement with the
experimental Hugoniots, not only for Ta, but also Al, Cu, Mo and W. Nevertheless, it seems unlikely that this simply model
will work for lower symmetry systems such as hcp-Fe \cite{wass} or for elastic constants. For example, the c/a varies
with temperature in hcp-Fe, but the Wang {\it et al.} model would not allow for this.

We summarize the zero pressure 300K equation of state parameters in Table \ref{table:p0params}. The equation of state gives $V$(0,T=300 K)=124.489 bohr$^3$, 2\% higher than the experimental value. Another way of looking at the discrepancy, is that the computed pressure at the room temperature experimental volume 121.8 bohr$^3$ is 4.1 GPa rather than 0. Table \ref{table:p0params} also shows the equation of state parameters computed at the experimental volume. The main discrepancy is the thermal expansivity which is 35\% too high, though this is reduced by comparing at the experimental volume. The Gruneisen parameter is similarly high. The origin of this discrepancy is unknown, as discussed above, although the thermal expansivity (and thus $\gamma$) are known to be very sensitive. Apparently this inaccuracy must decrease with increasing pressure, since our Hugoniot agrees well with experiment, up to temperatures of almost 10,000K. Perhaps our potential surface is not modeled accurately enough at small displacements and low pressures due to the very small energy differences involved in that regime.

\section{Conclusions}

We have studied the static and thermal equation of state of Ta from first
principles calculations. An electronic topological phase transition is found
around 200~GPa. Three different forms of thermal equation of state is
provided as: Vinet equation of state with temperature dependent equilibrium
quantities, simple linear temperature dependent average thermal pressure,
and a global fit to the Helmholtz free energy F(V,T). The simple equation of state $P_{th}=a T$
works quite well, but more accuracy and insites into higher order thermoelastic parameters were obtained from the global fit in $V$ and $T$. We find that $\alpha K_T$ is quite constant, as has been seen in experiments for a wide range of
materials above the Debye temperatures \cite{anderson} and has been shown for simple pair potentials \cite{hardy}.
Electronic excitations contribute significantly to the heat capacity temperature dependence of $C_V$, and thus
to variations in the Gr\"{u}neisen parameter $\gamma$. We find good agreement with the experimental Hugoniot
and thermal expansivity, though the rapid increase in the thermal expansivity and heat capacity at high temperatures
remains unexplained.

\acknowledgments{
This work was supported by DOE ASCI/ASAP subcontract B341492 to
Caltech DOE W-7405-ENG-48. Thanks to D. Singh and H.  Krakauer for
use of their LAPW code. Computations were performed on the Cray SV1
at the Geophysical Laboratory, supported by NSF grant EAR-9975753
and the W.\ M.\ Keck Foundation.  We thank J.L. Martins,
S. Mukherjee, G. Steinle-Neumann, L. Stixrude, and E.  Wasserman for
helpful discussions.}

\clearpage 
\begin{table}[tbp]
\caption{Vinet parameters for Ta isotherms. The temperatures are odd numbers
since computations were done at multiples of $k_BT=$0.006 Ry. }
\label{table:vinet}%
\begin{tabular}{l|llll}
T (K) & V$_0(T)$ (au) & K$_0(T)$ (GPa) & K$_0^\prime(T)$ & E$_0$(T) (Ryd) \\ 
\hline
0.00 & 123.63 & 190.9 & 3.99 & -31252.3341 \\ 
947.32 & 126.83 & 163.6 & 4.31 & -31252.3665 \\ 
2052.53 & 131.38 & 138.2 & 4.54 & -31252.4388 \\ 
2999.85 & 135.53 & 120.5 & 4.69 & -31252.5119 \\ 
3947.17 & 140.26 & 103.8 & 4.83 & -31252.5925 \\ 
5052.37 & 147.30 & 83.1 & 5.05 & -31252.6950 \\ 
5999.69 & 155.44 & 64.5 & 5.30 & -31252.7899 \\ 
6947.01 & 167.39 & 45.0 & 5.64 & -31252.8917 \\ 
8052.22 & 192.24 & 22.4 & 6.25 & -31253.0210 \\ 
8999.54 & 230 & 9.2 & 6.90 & -31253.1413 \\ 
9946.86 & 289 & 3.1 & 7.53 & -31253.2678 \\ 
&  &  &  & 
\end{tabular}%
\end{table}

\begin{table}[tbp]
\caption{Global fit parameters for Ta in SI units (except where marked). Row
is for i and column for j in A$_{ij}$ of Eq.~\ref{eq:globaleos}. Note that
the $T=0$ parameters are not identical to those in Table \ref{table:vinet},
since these were determined from a global fit to all results ($T=0$ and $T\ne0
$) and the Table \ref{table:vinet} values were fit to $T=0$ only. There is
no practical difference for applications of these equations of state, within
the accuracy of the computations.}
\label{table:params}%
\begin{tabular}{l|llll}
& 0 & 1 & 2 & 3 \\ 
\hline
1 & 2.768 10$^{-22}$ & -3.295 10$^{6}$ & 6.852 10$^{33}$ & -1.751 10$^{63}$
\\ 
2 & 3.734 10$^{-27}$ & -639.6 & 2.259 10$^{31}$ & -4.524 10$^{58}$ \\ 
3 & -1.955 10$^{-31}$ & 0.0247 & -6.512 10$^{26}$ & -1.137 10$^{55}$ \\ 
V$_0$ & 123.52 au & 1.8304 10$^{-29}$ m$^3$ &  &  \\ 
K$_0$ & 186.7 GPa & 186.7 GPa &  &  \\ 
K$_0^\prime$ & 4.120 & 4.120 &  &  \\ 
E$_0$ & -31252.3333 Ryd & -6.81261488 10$^{-14}$ J &  &  \\ 
&  &  &  & 
\end{tabular}%
\end{table}

\begin{table}[tbp]
\caption{Thermal equation of state parameters at ambient conditions (300 K) (theoretical room temperature volume and experimental room temperature volume).}
\label{table:p0params}
\begin{tabular}{l|lll}
& theoretical volume & experimental volume & experiment \\ 
\hline
$V_0$ bohr$^3$ & 124.5 &(121.8)& 121.8 \cite{cynn} \\ 
$K_{T0}$ GPa & 180 &197& 194 \cite{katahara} \\ 
$K_{T0}^\prime$ & 4.2 &4.07& 3.4 \cite{cynn} 3.8 \cite{katahara} \\ 
$\alpha_0$ 10$^{-5}$K$^{-1}$ & 2.64 &2.38& 1.95 \cite{katahara} \\ 
$\gamma_0$ & 2.09 &2.02& 1.64 \cite{katahara} \\ 
$C_P/R$ & 3.03 &3.08& 3.04 \cite{grigoriev} \\ 
\end{tabular}
\end{table}

\newpage \onecolumn 

\begin{figure}[tbp]
\centerline{\epsfig{file=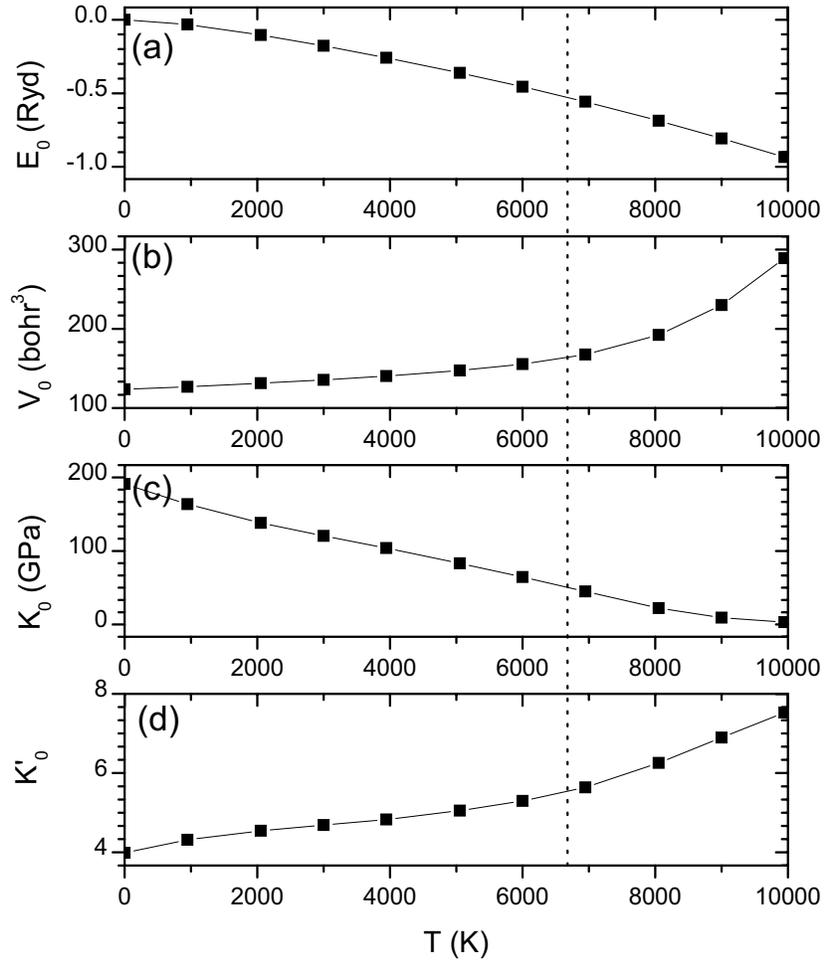,height=6in}}
\vspace{10pt} 
\caption{Vinet equation-of-state parameters as functions of temperature. (a)
The minimum energy with respect to the T=0 K value $E_0(T=0)=-31252.33412$%
~Ryd, (b) the equilibrium volume, $V_0$(T), (c) the bulk modulus $K_{T0}(T)$%
at $P=$0, and (d) the pressure derivative of bulk modulus at $P=$0. The
vertical line shows the position of last volume point. At higher
temperatures the parameters are fictive, but still govern the high pressure
equation of state.}
\label{fig:vinetvke}
\end{figure}

\begin{figure}[tbp]
\centerline{\epsfig{file=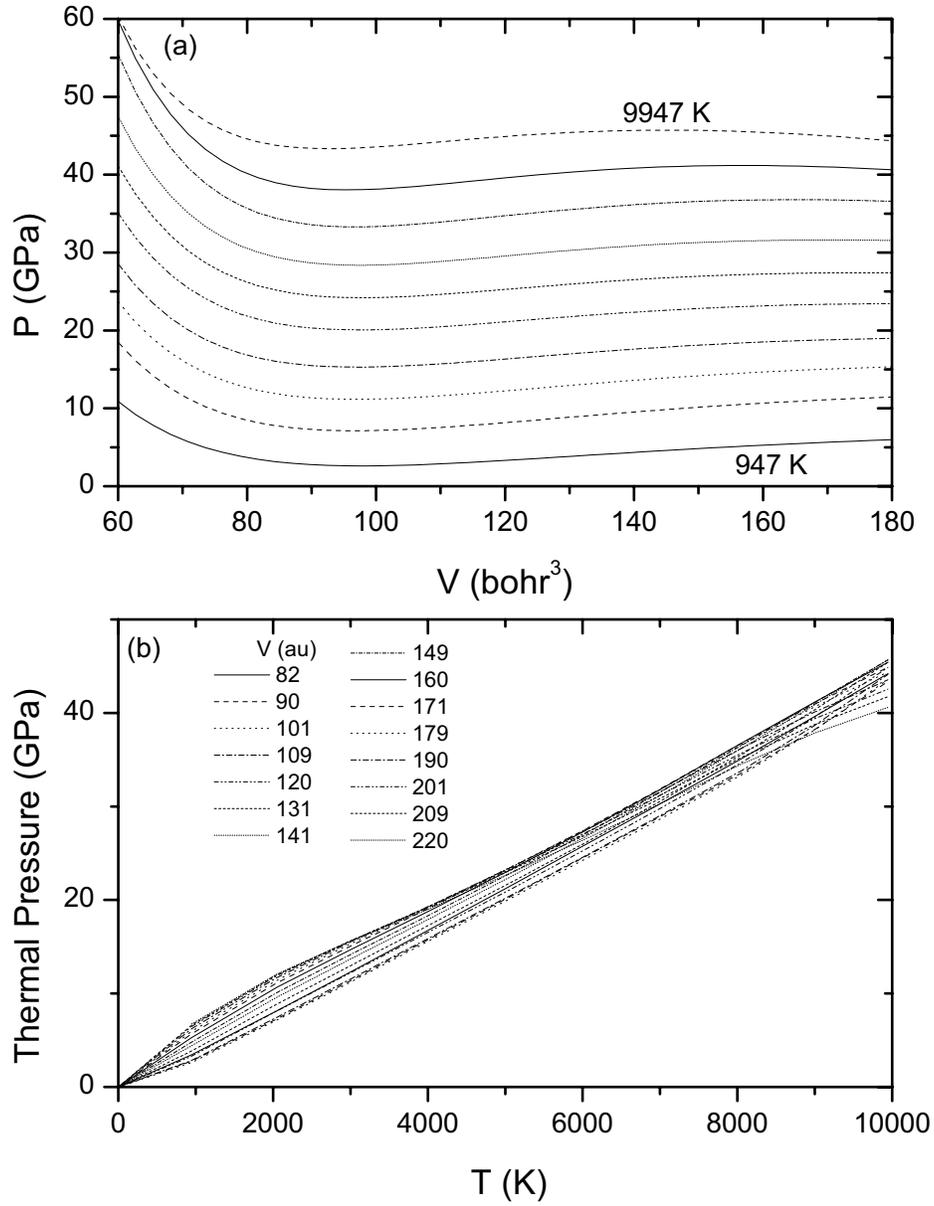,height=7in}}
\caption{The thermal pressure as a function of (a) volume at different
temperatures (shown in Table~\ref{table:vinet}) and (b) temperature from the
Vinet fits. }
\label{fig:pthermal}
\end{figure}

\begin{figure}[tbp]
\centerline{\epsfig{file=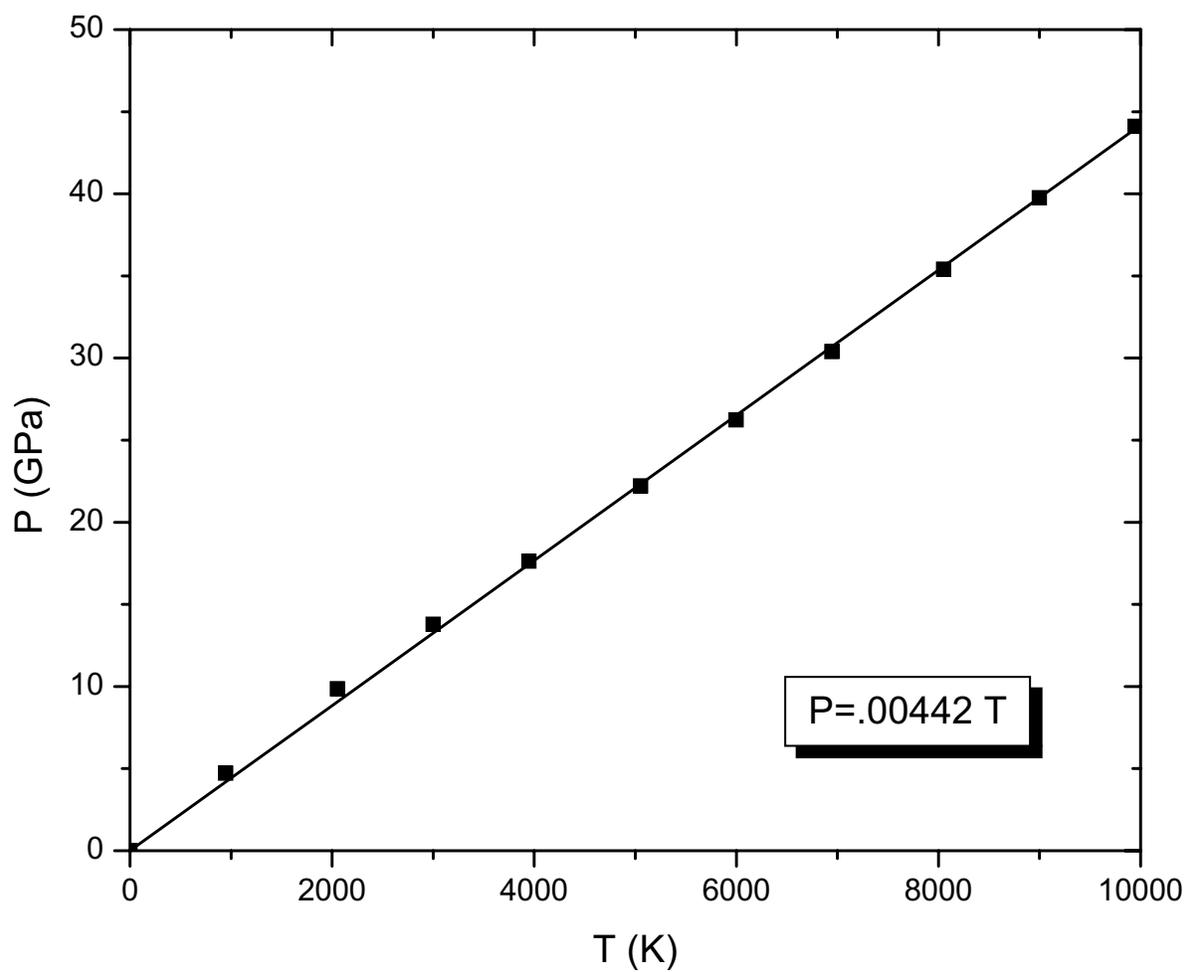,height=6in}}
\vspace{10pt} 
\caption{The thermal pressure averaged over volume as a function of
temperature. Solid line is a linear fit with slope equal to 0.00442~GPa/K.}
\label{fig:pavth}
\end{figure}

\begin{figure}[tbp]
\centerline{\epsfig{file=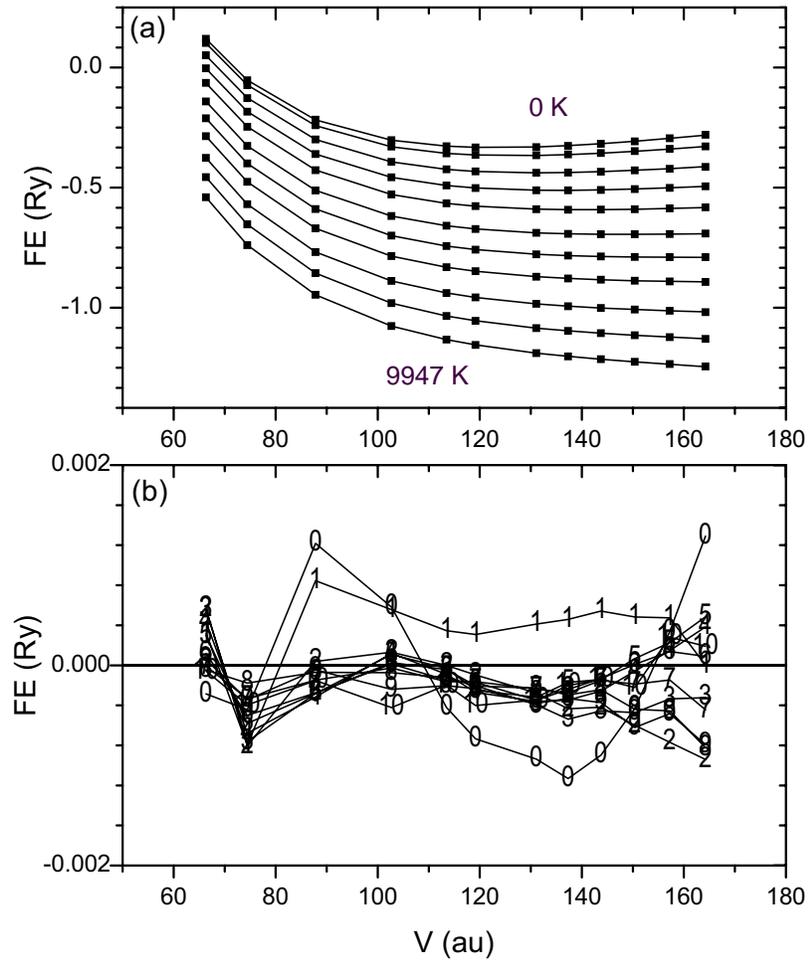,height=6in}}
\caption{Global fit to free energies. (a) Computed Helmholtz free energies
(symbols) and the result of the global fit (lines) using parameters from
Table \ref{table:params}. The top curve is for 0 K, and the bottom 9947 K.
See Table~\ref{table:vinet} for exact temperatures. (b) Residuals of fit.
Lines are labeled in order of temperature.}
\label{fig:globalf}
\end{figure}

\begin{figure}[tbp]
\centerline{\epsfig{file=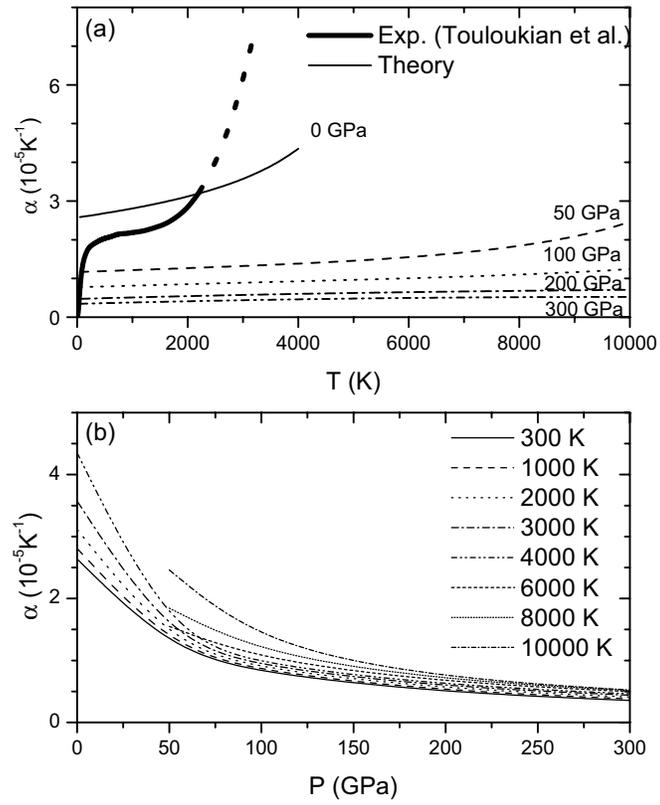,height=5in}}
\caption{The thermal expansivity as a function of (a) temperature and (b)
pressure. The wide line in (a) shows the experimental zero pressure thermal
expansivity (the dashed part is less well constrained) The experimental
thermal expansivity shows an anomalous increase at high temperatures. Theory
predicts the thermal expansivity to drop rapidly with pressure, and the
temperature dependence to decrease. }
\label{fig:alpha}
\end{figure}

\begin{figure}[tbp]
\centerline{\epsfig{file=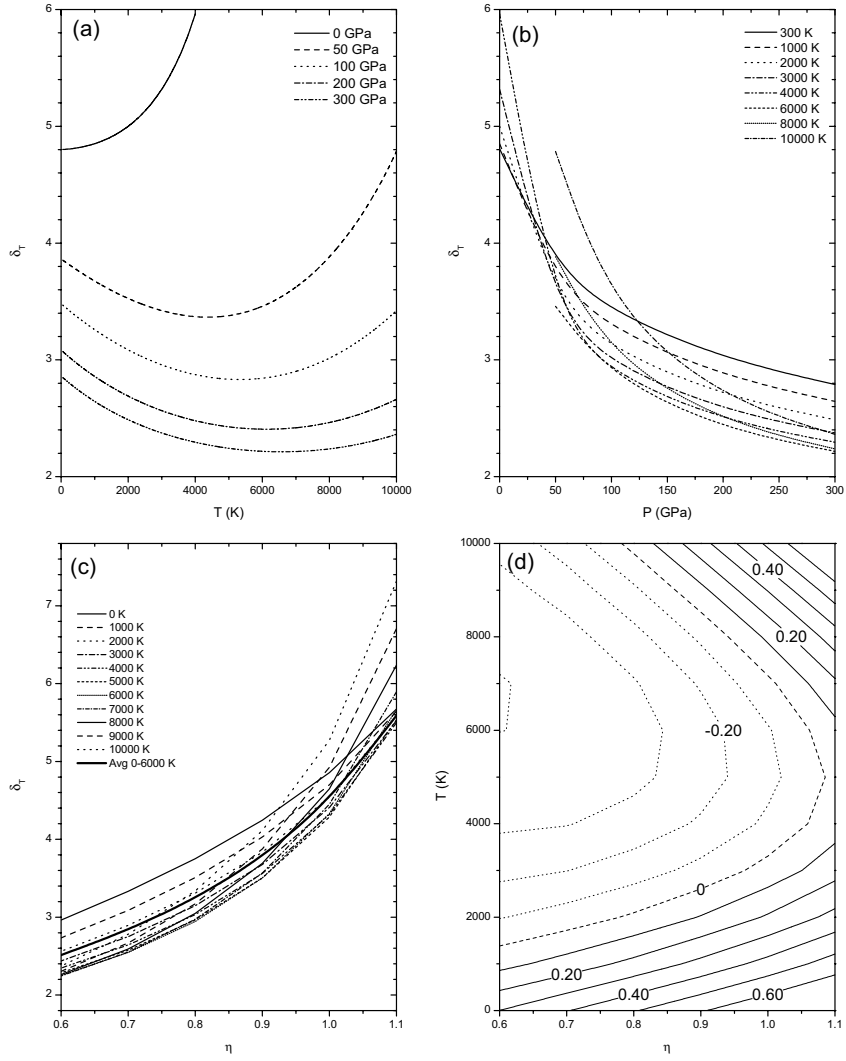,height=7in}}
\caption{The Anderson-Gr\"{u}neisen parameter $\protect\delta_T$ as a
function of (a) T, (b) P, and (c) $\protect\eta=V/V_0(T=0)$. The average of $%
\protect\delta _T$ from 0-6000 K is also shown as a thick line in (c). (d)
Contours of $\protect\delta _T - K^\prime$ show that this quantity is quite
small for 1000-2000 K, but increases at low and high temperatures.}
\label{fig:deltaT}
\end{figure}

\begin{figure}[tbp]
\centerline{\epsfig{file=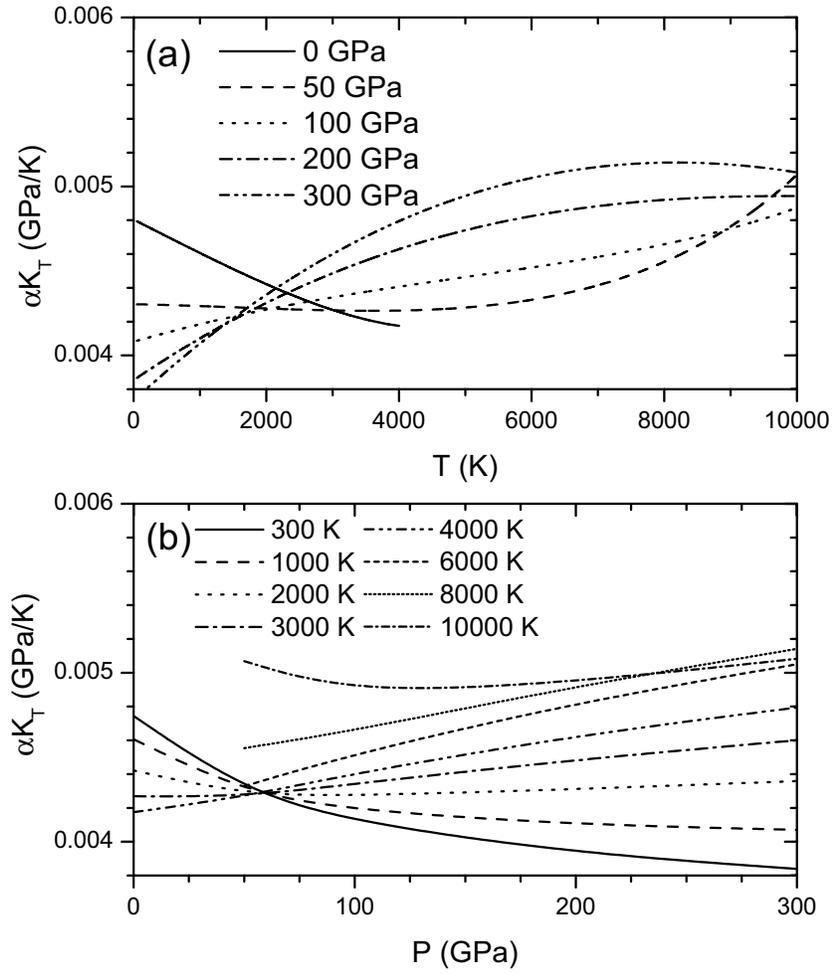,height=6in}}
\caption{The dependence of $\protect\alpha K_T$ on (a) temperature and (b)
pressure. The temperature dependence of $\protect\alpha K_T$ changes sign
with pressure, but the changes are small.}
\label{fig:akt}
\end{figure}

\begin{figure}[tbp]
\centerline{\epsfig{file=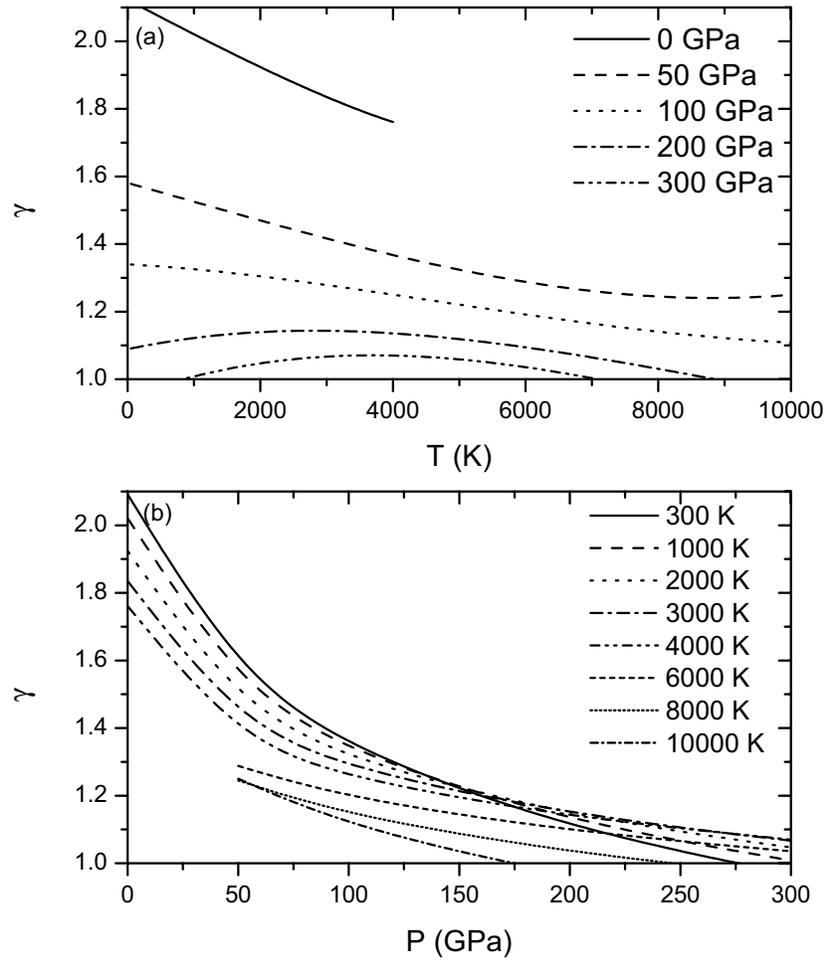,height=6in}}
\caption{Variation of the Gr\"{u}neisen parameter $\protect\gamma$ with (a)
temperature and (b) pressure. The temperature dependence is moderate.}
\label{fig:gamma}
\end{figure}

\begin{figure}[tbp]
\centerline{\epsfig{file=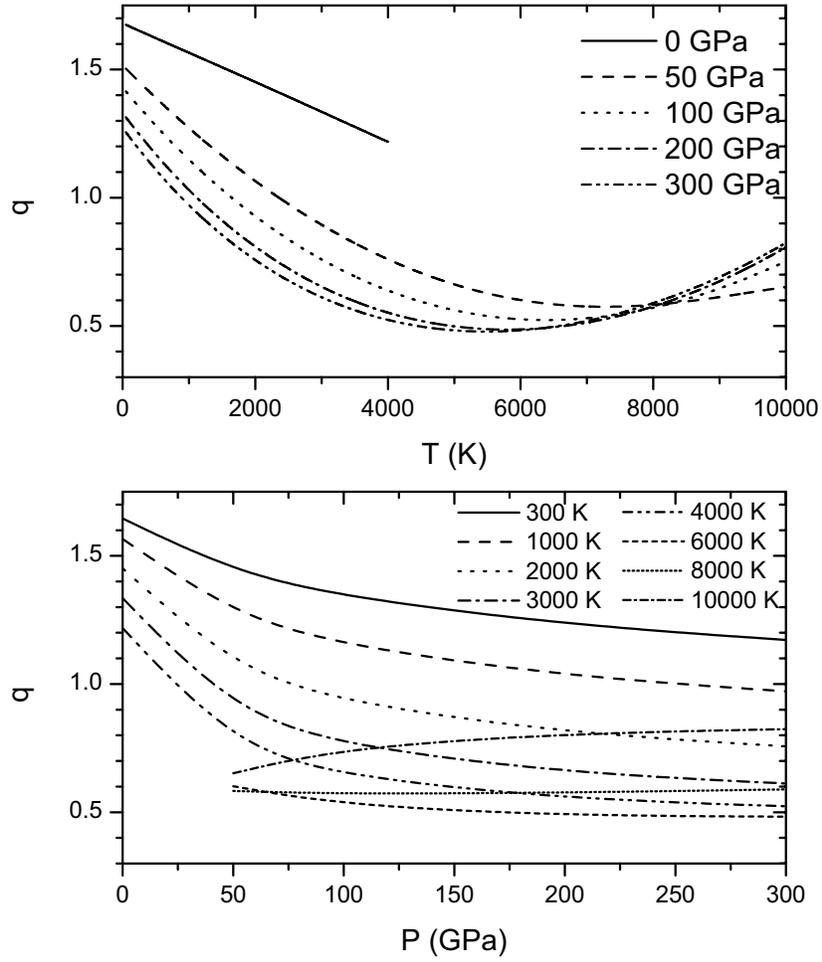,height=6in}}
\caption{Variation of q as a function of (a) temperature and (b) pressure.
The pressure dependence is small above 50 GPa, but the temperature
dependence is significant at all pressures.}
\label{fig:q}
\end{figure}

\begin{figure}[tbp]
\centerline{\epsfig{file=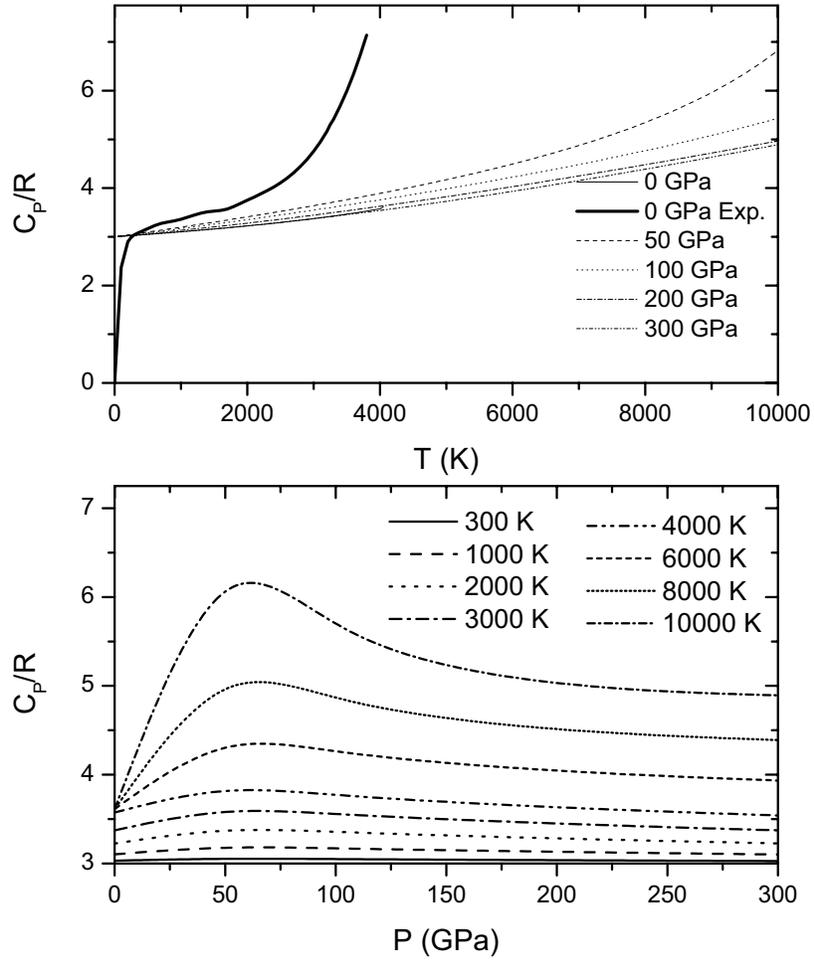,height=6in}}
\caption{Variation of the heat capacity with (a) temperature and (b)
pressure. The thick curve is the experimental heat capacity \protect\cite%
{grigoriev}, which shows an anomalous increase at high temperatures, similar
to the behavior of the thermal expansivity.}
\label{fig:heatcapacity}
\end{figure}

\begin{figure}[tbp]
\centerline{\epsfig{file=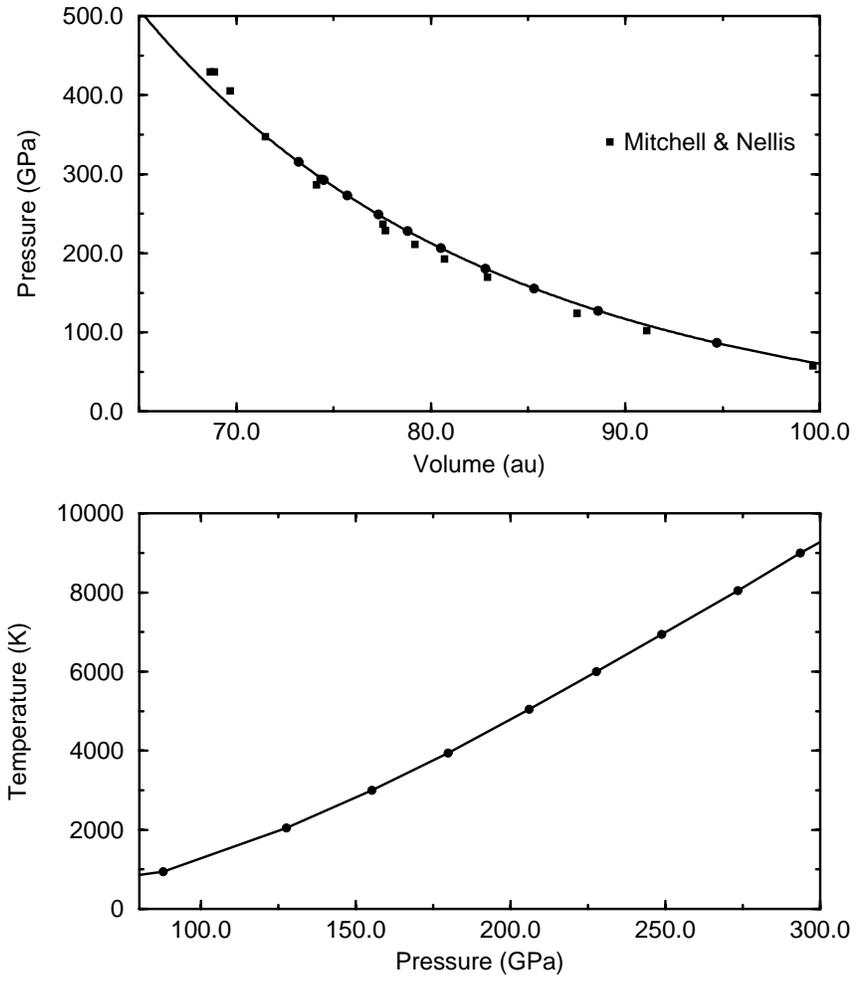,height=7in}}
\caption{a)The line shows the computed Hugoniot for Ta (computed at the points shown as circles) The squares are the shock-wave data %
\protect\cite{hug1}. b) Theoretical temperatures along the Hugoniot.}
\label{fig:hug}
\end{figure}

\end{document}